\documentclass[
  aps,
  pra,
  reprint,
  superscriptaddress,
  nofootinbib,
]{revtex4-2}
\setcitestyle{super,sort&compress,open={},close={}}
\usepackage[normalem]{ulem}
\usepackage{xcolor}
\usepackage{hyperref}
\usepackage{tikz}
\usetikzlibrary{shadings}
\usepackage{amsmath,amssymb}
\usepackage{mathrsfs}
\usepackage{lineno}
\usepackage[section]{placeins}
\usepackage{float}

\newcommand{\omegapS}{\omega_{p,\textrm{C}}}
\newcommand{\omegapC}{\omega_{p,\textrm{I}}}

\newcommand{\epsC}{\varepsilon_{r,\mathrm I}}

\newcommand{\epsS}{\varepsilon_{r,\mathrm C}}

\usepackage{graphicx,amsmath,amsfonts}% Include figure files
\usepackage{amssymb}

\newcommand{\omegarms}{\omega_\textrm{rms}}

\begin{document}

\title{Robust Fano-like Antiresonances in Large-Area Au-coated  Ag Nanoisland Ensembles}

\author{Valeria Nocerino}
\affiliation{Department of Electrical Engineering and Information Technology, Universit\`{a} degli Studi di Napoli Federico II, Via Claudio 21, 80125 Naples, Italy}
\affiliation{Centre for Advanced Metrology and Technological Services (CeSMA), 
Universit\`{a} degli Studi di Napoli Federico II, Corso Nicolangelo Protopisani, 80146 Naples, Italy}

\author{Francesca Di Carlo}
\affiliation{Department of Electrical Engineering and Information Technology, Universit\`{a} degli Studi di Napoli Federico II, Via Claudio 21, 80125 Naples, Italy}

\author{Bruno Miranda}
\affiliation{Institute of Applied Sciences and Intelligent Systems - National Research Council (ISASI-CNR), Via Pietro Castellino, 111, 80131 Naples, Italy}
\affiliation{Department of Electrical and Information Engineering, Polytechnic University of Bari, Via Orabona 4, 70125 Bari, Italy}

\author{Luca De Stefano}
\affiliation{Institute of Applied Sciences and Intelligent Systems - National Research Council (ISASI-CNR), Via Pietro Castellino, 111, 80131 Naples, Italy}

\author{Principia Dardano}
\affiliation{Institute of Applied Sciences and Intelligent Systems - National Research Council (ISASI-CNR), Via Pietro Castellino, 111, 80131 Naples, Italy}

\author{Carlo Forestiere}
\email[]{carlo.forestiere@unina.it}
\affiliation{Department of Electrical Engineering and Information Technology, Universit\`{a} degli Studi di Napoli Federico II, Via Claudio 21, 80125 Naples, Italy}

\begin{abstract}
We report robust Fano-like spectral profiles in substrate-supported Au-coated Ag nanoislands fabricated by solid-state dewetting followed by Au overgrowth. Combining electrostatic modal analysis, full-wave simulations, and linear transmission spectroscopy, we investigate the origin of these profiles, their tunability, and their robustness against morphological disorder. Increasing the Au coating thickness progressively red-shifts the low-energy resonance and transfers spectral weight to it from the high-energy resonance, whereas the Fano-like antiresonant dip undergoes a limited spectral displacement. The electrostatic modal analysis identifies this dip as the signature of destructive interference between superradiant and subradiant hybridized modes, and its weak spectral evolution is consistently reproduced by theory, simulations, and experiments. These results establish Au-coated Ag nanoislands as a scalable, lithography-free platform for engineering robust yet tunable Fano-like optical responses accessible through standard far-field spectroscopy.
\end{abstract}

\maketitle

%\section{Introduction}

Multimaterial plasmonic nanostructures provide a powerful route to engineer optical responses beyond those achievable in homogeneous metallic systems, because the composition and geometry of each constituent material can be tuned within the same nano-object \cite{haMulticomponentPlasmonicNanoparticles2019,sytwu_bimetallic_2019}. In bimetallic architectures, the coexistence of two distinct dielectric functions introduces additional degrees of freedom:  the plasmonic modes depend not only on the overall shape of the particle, but also on the dielectric contrast between the two metals and on the shape of the region occupied by each of them \cite{chakrabortyEvolutionColloidalPlasmonic2024,sahaElectronEnergyLoss2024}. This enables additional control over mode hybridization, interference, and emergent scattering profiles in compact plasmonic systems \cite{prodanHybridizationModelPlasmon2003, mamoGeometricDielectricModulation2025}.

Ag–Au nanostructures are the prototypical bimetallic platform, combining the complementary properties of the two metals \cite{heAgAuBimetallic2025,awiazRecentAdvancesAu2023,tuff_ion_2023}. Silver supports intense and relatively narrow localized surface plasmon resonances, but suffers from oxidation and chemical instability \cite{kangStabilizationSilverGold2018}. Gold is chemically robust and plasmonically active in the visible, though its interband transitions add optical losses there \cite{haMulticomponentPlasmonicNanoparticles2019,lozaSynthesisStructureProperties2020,ghoshchaudhuriCoreShellNanoparticles2012}. Combining them therefore exploits the strong response of Ag while gaining the stability and characteristic dispersion of Au.

Several Ag-Au architectures have been explored, including alloys \cite{bonviciniFormationRemovalAlloyed2022, ametrano2024structural}, heterodimers \cite{luSwitchingPlasmonicFano2019}, nanoparticle assemblies \cite{zorattiSimpleOneStepSynthesis2025}, Janus particles \cite{weller_gap-dependent_2016}, oligomer clusters \cite{aldufeeryDisentanglingBrightDark2026}, and core--shell nanoparticles \cite{heAgAuBimetallic2025,srnova-sloufova_coreshell_2000,rodriguez-gonzalez_multishell_2005,zhang_surface_2015}. In each of these systems, coupling between Ag- and Au-related plasmonic modes offers a further handle on the optical response through mode hybridization and interference \cite{hangPlasmonicSilverGold2024,ghoshchaudhuriCoreShellNanoparticles2012,muhammadDualOptimizationShell2025,liHighlyHomogeneousBimetallic2023}.

A direct consequence of such coupling among plasmon modes is the possible  emergence of asymmetric (\textit{i.e.}, \textit{Fano-like}) spectral profiles
\cite{lukyanchukFanoResonancePlasmonic2010,Miroshnichenko2010, zebFanoResonanceStrongcoupling2022, wangTunableFanoResonance2018}. In plasmonic systems, these features generally arise from destructive interference between resonant channels with different radiative strengths. A broad, strongly radiative mode, often referred to as a \textit{superradiant} mode, can interfere with a narrower and weakly radiative \textit{subradiant} mode, producing an asymmetric spectral profile and a pronounced antiresonant minimum \cite{forestiereTheoryCoupledPlasmon2013}. Although Fano-like resonances have been reported in several bimetallic plasmonic architectures, their observation often relies on finely engineered nanostructures \cite{bachelierFanoProfilesInduced2008,Lombardi2016,zhuMultipleFanoResonances2026, alvesEnhancedSurfaceFields2024, barelliColorRoutingCrosspolarized2020}. This is because the interference responsible for the asymmetric line shape is highly sensitive to structural symmetry, excitation polarization, incidence direction, interparticle spacing, and relative particle orientation \cite{leeControlledAssemblyPlasmonic2021,changPlasmonicFanoSwitch2012}. Consequently, in ensemble-averaged measurements, where these parameters are inevitably distributed over many nanostructures, the Fano-like signature is often broadened or partially washed out by the dominant plasmonic background. For this reason, polarization-resolved \cite{attiaouiPolarizationTunedFanoResonances2023}, angle-resolved \cite{liPhaseControlledFanoResonances2026}, near-field \cite{flauraud_mode_2017}, or nonlinear spectroscopies are frequently required to isolate the interference contribution \cite{bachelierFanoProfilesInduced2008,Lombardi2016,zhuMultipleFanoResonances2026,alvesEnhancedSurfaceFields2024}.

Within this context, Ag--Au core--shell nanoparticles represent a particularly appealing bimetallic platform, which may support Fano-like spectral responses \cite{pena-rodriguezAuAgCore2011,pena-rodriguezEnhancedFanoResonance2011,wangPlasmonicCoreShell2022}. Since plasmon hybridization occurs within a single nanostructure \cite{mukherjeeFanoshellsNanoparticlesBuiltin2010}, these systems do not rely on precisely controlled interparticle coupling, as instead required in many assembled or oligomeric architectures.  Despite these advantages, conventional core--shell nanoparticles are most commonly realized as colloidal systems in solution, whereas many sensing and integrated-photonics applications require nanostructures directly supported on a substrate. In this respect, coated nanoislands (NIs) offer a complementary and technologically attractive route \cite{chungNanoislandsPlasmonicMaterials2019,kvitekPreparationAlloyedCoreshell2020}: they retain the hybridization mechanism of core--shell systems, while being naturally compatible with planar substrates, large-area fabrication, and standard far-field optical characterization. 

Despite this potential, Fano-like interference in substrate-supported bimetallic-coated NIs fabricated by scalable, lithography-free routes remains largely unexplored, both in its physical origin and in its tunability. It is equally unclear how the morphological disorder inherent to these fabrication routes affects plasmon hybridization, modal interference, and the resulting far-field optical response.

Motivated by these open questions, we report Fano-like spectral profiles in substrate-supported Au-coated Ag nanoislands (AgNIs) and investigate their origin, tunability, and robustness against morphological disorder.

Specifically, we develop an electrostatic theoretical framework to study resonances and interference effects in a generic bimetallic-coated NI.  The analysis identifies the hybridized plasmonic modes supported by the coated NI, their tunability in terms of the plasma-frequency mismatch between the two metals and of the geometry, and shows how modal interference can suppress the total induced dipole moment. This dipolar cancellation produces an antiresonant minimum, giving rise to a Fano-like asymmetric profile in the optical response. Guided by this theoretical framework, we fabricate Au-coated AgNIs and characterize their optical response by conventional linear extinction spectroscopy. The measured spectra reveal two main resonant features separated by an antiresonant dip, consistent with interference between hybridized plasmonic modes. Full-wave numerical simulations using realistic dielectric functions for Ag and Au further support this interpretation. The combined theoretical, experimental, and numerical analysis reveals that the hybridized plasmonic resonances shift significantly with increasing Au coating thickness, whereas the antiresonant minimum remains comparatively stable. These results demonstrate that Au-coated AgNIs provide a simple and scalable platform for realizing robust Fano-like optical responses, even in structurally disordered ensembles.

This combination of a tunable resonance with a fabrication-insensitive antiresonance is promising for wafer-scale plasmonic biosensing \cite{waitkusGoldNanoparticleEnabled2023}, where reliable spectral reproducibility is often more important than maximizing peak sharpness.

The paper is organized as follows. 
Section~\ref{sec:EStheory} introduces the electrostatic modal framework used to describe the resonances of bimetallic-coated NIs and to identify the destructive modal interference responsible for the Fano-like antiresonance. 
Section~\ref{sec:Results} applies this framework to Au-coated AgNIs, combining numerical simulations, morphological characterization, and optical extinction measurements to elucidate the evolution of the resonant and antiresonant features with Au shell thickness. 
Finally, Section~\ref{sec:Conclusion} summarizes the main findings and discusses the implications of robust Fano-like interference in scalable, disordered Au-coated Ag NI ensembles.

\section{Electrostatic resonances of bimetallic nanoislands}

\label{sec:EStheory}

\begin{figure}
    \centering
    \includegraphics[width=0.8\linewidth]{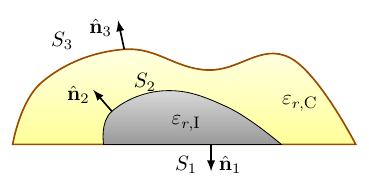}
        \caption{Schematic of an arbitrarily shaped homogeneous NI partially coated by a second homogeneous material. The NI, coating, and surrounding medium occupy the domains \(V_1\), \(V_2\), and \(V_3\), with relative permittivities \(\epsC\), \(\varepsilon_{r,\mathrm{C}}\), and \(1\), respectively. The NI--coating interface is denoted by \(S_2\), while \(S_1\) and \(S_3\) indicate the portions of the NI and coating surfaces exposed to the exterior medium. The unit normals \(\hat{\mathbf{n}}_1\), \(\hat{\mathbf{n}}_2\), and \(\hat{\mathbf{n}}_3\) are associated with \(S_1\), \(S_2\), and \(S_3\), respectively.}
    \label{fig:Sketch_NI}
\end{figure}

We first analyze a generic subwavelength bimetallic-coated nanoisland (NI) within the quasi-electrostatic limit, with the geometrical quantities defined in Figure~\ref{fig:Sketch_NI}. We formulate the driven scattering problem under a time-harmonic incident electric field, $\mathbf{e}_{\rm inc}(\mathbf{r},t)= \text{Re} \left\{ \mathbf{E}_{\mathrm{inc}}(\mathbf{r})\,e^{-i\omega t}\right\}$, at angular frequency $\omega$. The NI and its coating are assumed to be linear, homogeneous, isotropic, time-invariant, and temporally dispersive media, with relative permittivities \(\epsC(\omega)\) and \(\epsS(\omega)\), respectively. The exterior medium is vacuum, so that \(\varepsilon_{r,3}=1\). 
Let $a$ denote the radius of the smallest circumscribing sphere that encloses the whole particle. Throughout this section we consider the subwavelength regime, \(k_0 a\ll 1\), where $k_0 = \omega / c$  is the free-space wavenumber and $c$ is the speed of light in vacuum. In this limit, the electromagnetic response is well described by the quasi-electrostatic approximation, and radiative effects may be neglected.

%By recasting the corresponding boundary-integral equation as an eigenvalue problem, we identify the plasmonic resonances supported by the coated structure and explicitly separate the effects of geometry from the additional material degree of freedom associated with the distinct dielectric responses of the two metals. We then determine how these eigenmodes couple to the incident field and show that their contributions to the induced dipole moment can interfere destructively. This modal cancellation suppresses the dipolar response, giving rise to an antiresonant minimum and, consequently, to a Fano-like scattering profile.

In the quasi-electrostatic limit, the solution of the electromagnetic scattering problem can be expressed in terms of surface charge densities \(\sigma_1\), \(\sigma_2\), and \(\sigma_3\) distributed over the NI--environment $S_1$,  NI--coating $S_2$, and coating--environment interfaces $S_3$, respectively. Following the boundary-integral framework of Refs. \cite{mayergoyz_electrostatic_2005,fredkin_resonant_2003,mayergoyz_plasmon_2013}, these densities satisfy the driven integral equation
\begin{equation}
\left[ \epsS (\omega)
\mathcal{A} + \epsC (\omega) \,
\mathcal{B}
+ \mathcal{C} \right]
\boldsymbol{\sigma} = \mathbf{f}_\mathrm{inc} (\omega)
, 
\label{eq:DrivenProblem}
\end{equation}
where $\boldsymbol{\sigma}$ is the vector of  charge density unknowns
$
\boldsymbol{\sigma} = 
\left[
\sigma_1,
\sigma_2,
\sigma_3   
\right]^\intercal
$, ${\mathcal{A}}$, ${\mathcal{B}}$, and ${\mathcal{C}}$ are
surface-integral block-operators which depend neither on the material dispersion nor on frequency, but only on the geometry of the system defined
by Eqs. \eqref{eq:OperatorA},\eqref{eq:OperatorB},\eqref{eq:OperatorC} of the Method section \ref{sec:MethodEigProb}
 while the right-hand side is defined by Eq. \eqref{eq:incident_source}.

We assume a Drude dispersion relation for both the island and the coating:
\begin{equation}
\varepsilon_{r\text{I}} (\omega) \approx 1 - \frac{\omegapC^2}{\omega ( \omega + i \nu_\mathrm{I})} ; \quad     \varepsilon_{r\text{C}} (\omega) \approx 
      1 - \frac{\omegapS^2}{\omega ( \omega + i \nu_\mathrm{C})}. 
\label{eq:DrudeDispersion}
\end{equation}
Here, \(\omegapC\) and \(\omegapS\) denote the plasma frequencies of the internal NI and the coating, respectively, while \(\nu_{\mathrm{I}}\) and \(\nu_{\mathrm{C}}\) are the corresponding damping rates.

By setting the excitation term on the right-hand side of Eq.~\eqref{eq:DrivenProblem} to zero, substituting Eqs. \eqref{eq:DrudeDispersion} and assuming the lossless limit by setting $\nu_{\mathrm{I}}=\nu_{\mathrm{C}}=0$, we obtain the eigenvalue problem
\begin{equation}
\left(\mathcal E+\delta \mathcal F\right)
\boldsymbol{\sigma}^{(k)}
=
\tilde{\omega}_k^2
\boldsymbol{\sigma}^{(k)} ,
\label{eq:main_eigenproblem}
\end{equation}
where
$
\boldsymbol{\sigma}^{(k)} = 
\left[
\sigma_1^{(k)},
\sigma_2^{(k)},
\sigma_3^{(k)}   
\right]^\intercal
$ 
collects the surface-charge densities associated with the \(k\)-th eigenmode on the three interfaces. The corresponding normalized resonance frequency is
\(\tilde{\omega}_k=\omega_k/\omegarms\), where
\begin{equation}
\omegarms= \sqrt{\frac{\omegapC^2+\omegapS^2}{2}}.
\label{eq:root_mean_square}
\end{equation}
is the root mean square of the plasma frequencies of the two metals. The dimensional resonance frequencies are then recovered through \(\omega_k=\tilde{\omega}_k \, \omegarms\). 
The surface-integral block operators \(\mathcal{E}\) and \(\mathcal{F}\), which depend exclusively on the geometry of the coated NI, are defined in Eqs.~\eqref{eq:IntegralOperatorE} and \eqref{eq:IntegralOperatorF} of the Methods section. The material contrast between the two metallic regions is instead captured by the dimensionless plasma-frequency mismatch
\begin{equation}
\delta=
\frac{\omegapS^2-\omegapC^2}
{\omegapS^2+\omegapC^2}.
\label{eq:main_delta}
\end{equation}
Positive values of \(\delta\) correspond to a coating whose plasma frequency exceeds that of the internal NI, whereas negative values indicate the opposite configuration. For the Au-coated AgNIs considered here, \(\delta<0\), since the plasma frequency of Ag is larger than that of Au.  The eigenvalue formulation therefore cleanly separates the influence of geometry, encoded in the integral operators \(\mathcal{E}\) and \(\mathcal{F}\), from that of material composition, encoded by the single contrast parameter \(\delta\). We refer the reader to  Sec. \ref{sec:MethodEigProb} for a detailed derivation of the eigenvalue problem in Eq. \eqref{eq:main_eigenproblem}.

The normalized resonances are scale invariant \cite{mayergoyz_electrostatic_2005}: for a fixed overall shape, \(\tilde{\omega}_k\) depends on dimensionless geometrical parameters, such as the ratio between the coating thickness and the NI radius, and on \(\delta\), but not on the absolute NI size. Equation~\eqref{eq:main_eigenproblem} therefore identifies the plasmonic resonance frequencies and surface-charge modes supported by the coated NI, while providing a compact description of how the choice of metals and the coating geometry modify the modal spectrum.

The electric dipole moment $\mathbf{p}$ induced in  the coated NI by the external electric field at angular frequency $\omega$, corresponding to the normalized frequency \(\tilde{\omega}=\omega/\omegarms\), can be expanded in terms of the dipole moments $\left\{ \mathbf{p}^{(k)} \right\}$, defined in Eq. \eqref{eq:modal_dipole}, and associated with the surface charge density modes $\boldsymbol{\sigma}^{(k)}$. For a constant incident field polarized along the \(x\)-direction, i.e. $\mathbf{E}_{inc} = \hat{\mathbf{x}}$, the induced dipole moment along  $\hat{\mathbf{x}}$, $p_x = \mathbf{p} \cdot \hat{\mathbf{x}}$, is
\begin{equation}
p_x(\tilde{\omega})
=
\sum_{k} 
\frac{\alpha_k }
{\tilde{\omega}_k^2-\tilde{\omega}^2} p_x^{(k)},
\label{eq:main_modal_dipole}
\end{equation}
where \(\alpha_k\) is the excitation coefficient of the \(k\)-th mode defined in Eq. \eqref{eq:alpha_k} and $p_x^{(k)} = \mathbf{p}^{(k)} \cdot \hat{\mathbf{x}}$ is the  dipole moment of the $k$-th plasmonic mode projected along the incident polarization direction.  We refer the reader to  Sec. \ref{sec:coupling} for a detailed derivation of Eq. \eqref{eq:main_modal_dipole}. Within the electric-dipole approximation and for a homogeneous vacuum background, the corresponding scattered power is
\begin{equation}
P_{\mathrm{sca}}=\frac{\omega^4}{12 \pi \varepsilon_0 c^3}|\mathbf{p}|^2=\frac{\omega^4}{12 \pi \varepsilon_0 c^3} \sum_{t \in\{x, y, z\}}\left|p_t\right|^2.
\label{eq:Psca}
\end{equation}

It is convenient to introduce the modal strength $ s_k=\alpha_k p_x^{(k)}$, which quantifies the combined excitation efficiency and dipolar strength of the \(k\)-th mode and, within the present electrostatic formulation, is independent of frequency.  Equation~\eqref{eq:main_modal_dipole} shows that the measured optical response is not determined by the eigenfrequencies alone, but also by how efficiently each mode is excited and how strongly it radiates.

An \textit{antiresonance} occurs when these contributions interfere destructively and cancel the induced dipole moment at the normalized antiresonance frequency $\tilde{\omega}_{\mathrm{ar}}$. In the electrostatic limit, the antiresonance condition is therefore \cite{forestiereTheoryCoupledPlasmon2013,forestiere_cloaking_2014}
\begin{equation}
p_x(\tilde{\omega}_{\mathrm{ar}})=0.
\label{eq:main_zero}
\end{equation}Eq.~\eqref{eq:main_zero} describes an exact cancellation of the dipolar response for an ideal lossless system. When \(p_x\) is the only nonvanishing dipole component, Eq.~\eqref{eq:Psca} then predicts a complete suppression of the dipolar scattered power. In realistic Au-coated AgNIs, material absorption and radiative damping generally prevent an exact zero at a real frequency, replacing it with a finite antiresonant minimum. Moreover, even when the dipolar contribution is strongly suppressed, higher-order multipoles may retain a nonzero scattering response. The resulting minimum appears in the far-field spectrum as an asymmetric Fano-like dip produced by destructive interference between plasmonic modes with different radiative strengths.

When the optical response is dominated by two modes, labelled \(1\) and \(2\), having normalized resonance frequencies $\tilde{\omega}_1$ and $\tilde{\omega}_2$, Eq.~\eqref{eq:main_modal_dipole} reduces to
\begin{equation}
p_x(\tilde{\omega})
\simeq
\frac{s_1}{\tilde{\omega}_1^2-\tilde{\omega}^2}
+
\frac{s_2}{\tilde{\omega}_2^2-\tilde{\omega}^2},
\label{eq:main_two_mode}
\end{equation}
where \(s_1\) and \(s_2\) are the corresponding modal strengths. The antiresonance occurs at the frequency for which the two terms in Eq.~\eqref{eq:main_two_mode} cancel. Imposing Eq. \eqref{eq:main_zero} yields
\begin{equation}
\tilde{\omega}_{\mathrm{ar}}
=
\sqrt{
\frac{
s_1\tilde{\omega}_2^2
+
s_2\tilde{\omega}_1^2
}{
s_1+s_2
}}
.
\label{eq:two_mode_antiresonance}
\end{equation}

The antiresonance frequency is governed by the relative modal strengths and is therefore not generally centered between the two resonances. Assuming \(s_1>s_2\), mode~1 provides the stronger, superradiant contribution, whereas mode~2 corresponds to the weaker, subradiant contribution. Consequently, the antiresonance is shifted toward the weaker mode and lies closer to its resonance frequency, \(\tilde{\omega}_2\). This two-mode picture captures the essential physics of the Au-coated AgNIs: a broad radiative mode interferes with higher-order, weakly radiative modes, producing an asymmetric line shape and an antiresonant minimum between the main resonant features.

The electrostatic model therefore provides a simple physical interpretation of the bimetallic NI response. The plasma-frequency mismatch controls the hybridized modal spectrum, the ratio between the coating thickness and the NI radius tunes the relative position and strength of the modes, and destructive interference between dipolar contributions gives rise to a Fano-like antiresonance. Although the Drude model neglects the visible interband transitions of Au, it captures the fundamental role of direct metal–metal coupling in determining the hybridized modal spectrum and the associated Fano-like interference. The influence of realistic dispersive dielectric functions is subsequently investigated through full-wave electromagnetic simulations. Moreover, as shown in Sec.~\ref{sec:substrate}, the same quasi-electrostatic framework can be extended to substrate-supported particles by replacing the free-space Green function with the corresponding half-space Green function.

\section{Results}
\label{sec:Results}

\subsection{Numerical Results}

\begin{figure*}
\centering
\includegraphics[width=0.85\linewidth]{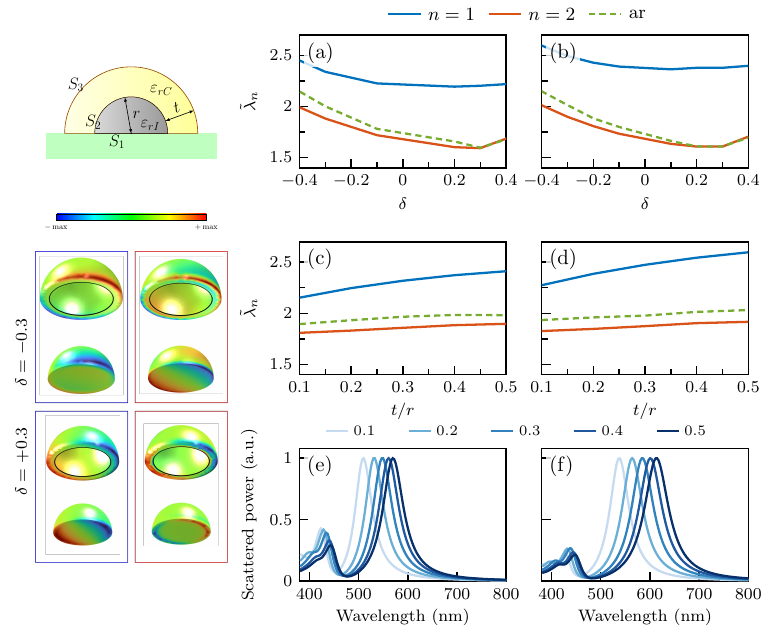}
\caption{Coated hemispherical NI with thickness $t$ and radius $r$. (Left) Surface-charge density distributions of the first two non-degenerate dipolar eigenmodes with non-zero net dipole moment for two values of the plasma-frequency mismatch parameter: $\delta=-0.3$ (top row) and $\delta=+0.3$ (bottom row). The eigenmode distributions are arranged in two columns: the left column (blue frame) corresponds to the fundamental dipolar mode ($n=1$), whereas the right column (red frame) corresponds to the higher-order dipolar mode ($n=2$). Normalized resonant wavelengths $\tilde{\lambda}_n$ of the first two non-degenerate electric dipole modes and the corresponding first antiresonance ($\tilde{\lambda}_{ar}$) as functions of the plasma-frequency mismatch $\delta$ for $t/r=0.3$: (a) without substrate and (b) with substrate. Normalized resonant wavelengths of the first two dipolar modes and the corresponding first antiresonance ($\tilde{\lambda}_{ar}$) as functions of the ratio $t/r$ for the fixed value $\delta=-0.2739$, corresponding to Au-coated AgNIs: (c) without substrate and (d) with substrate. Scattered-power spectra as functions of the ratio $t/r$ for the same conditions: (e) without substrate and (f) with substrate.}
\label{fig:FigTheory}
\end{figure*}

Here we study in detail the coated hemisphere shown in the top-left of Figure \ref{fig:FigTheory}. The coated hemisphere has internal radius $r$ and a coating thickness $t$.  As already mentioned, in the electrostatic limit, plasmon resonances are scale invariant and therefore remain unchanged under a uniform scaling of all geometrical dimensions. For this geometry, they only depend on the ratio between $t$ and $r$, but not on their individual values. 

We compute the first two non-degenerate modes with non-vanishing dipole moments. The remaining modes have much smaller dipole moments and are not considered since they play a minor role. Then, we analyze the shift of the resonance frequencies of these two modes as a function of the plasma-frequency mismatch parameter $\delta$. To systematically assess the role of plasma frequency mismatch, $\delta$ is varied in the interval $-0.4 \leq \delta \leq 0.4$, which spans physically relevant contrasts between the plasma frequencies of the NI and coating materials. The case $\delta = 0$ corresponds to the symmetric limit in which the two materials have identical plasma frequencies, and the particle is effectively homogeneous, whereas nonzero values of $\delta$ progressively increase the degree of asymmetry.
%%%%%%%
%%%%%%%
Figure~\ref{fig:FigTheory} (a) shows the first two dipolar resonance wavelengths normalized to $\lambda_\Sigma = 2 \pi c / \omegarms$ together with the corresponding first antiresonance, denoted by $\tilde{\lambda}_{ar}$, as functions of $\delta$, both in the absence (a) and in the presence of a substrate (b). In both cases, increasing $\delta$ shifts the dipolar resonances toward shorter wavelengths. The effect is mode dependent: the higher-order dipolar mode ($n=2$) exhibits a significantly larger spectral shift than the fundamental mode ($n=1$), which varies only weakly over the investigated range. The sensitivity to the plasma-frequency mismatch is greater for negative values of $\delta$ and gradually decreases for positive values. The first antiresonance ($\tilde{\lambda}_{ar}$) follows the same overall trend, shifting toward shorter wavelengths as $\delta$ increases while remaining located between the two dipolar resonances, consistent with its origin as the destructive-interference minimum associated with the Fano-like response. The substrate induces an overall redshift of both resonances, consistent with the increase in the effective refractive index of the surrounding medium, while preserving the same qualitative dependence on $\delta$. As a result, $\delta$ primarily controls the spectral position of the dipolar resonances, whereas the substrate mainly provides a nearly uniform spectral offset.

We now consider the specific Au-coated AgNI system investigated experimentally in the following sections. The effective Drude plasma frequencies are taken as $\omega_{p,\mathrm{Au}} = 6.790 \times 10^{15}\,\mathrm{rad/s}$ and $\omega_{p,\mathrm{Ag}} = 8.994 \times 10^{15}\,\mathrm{rad/s}$.  These values were determined through an inverse fitting procedure. For each metal, the plasma frequency was adjusted so that the peak of the power scattered by a small sphere described by the Drude model coincides with that obtained using the optical constants of Raki\'{c} \textit{et al.}, based on the Brendel--Bormann model \cite{rakicOpticalPropertiesMetallic1998}.

Their root mean square is $\omegarms = 7.9686 \times 10^{15} $. Thus, the normalized plasma frequencies are $\tilde{\omega}_{p,\mathrm{Au}} = 0.85$ and $\tilde{\omega}_{p,\mathrm{Ag}} = 1.13$, and the  mismatch parameter is $\delta = -0.2739$. The corresponding Drude damping rates are
$\nu_{\mathrm{Au}}=0.0433\,\omega_{p,\mathrm{Au}}$
and
$\nu_{\mathrm{Ag}}=0.0363\,\omega_{p,\mathrm{Ag}}$. The damping rates $\nu_{\mathrm{Ag}}$ and $\nu_{\mathrm{Au}}$
were estimated from the full widths at half maximum of the dipolar scattering resonances calculated using Mie theory for homogeneous Au and Ag nanospheres, using the optical constants reported in \cite{rakicOpticalPropertiesMetallic1998}.

These values were subsequently employed in the Drude model for the calculation of the scattered-power spectra presented in Figure~\ref{fig:FigTheory}(e,f). 

For this fixed value of $\delta$, we investigate the role of geometry by varying the ratio between the coating thickness and NI radius $t/r$. The ratio $t/r$ is varied within the interval $0.1 \leq t/r \leq 0.5$ to investigate the effect of geometry on the modal structure. This range is chosen to span physically meaningful experimentally accessible configurations, from relatively thin coating to cases in which the coating thickness is comparable to the NI radius. As a result, the analysis captures the progressive transition from weak to strong geometrical influence of the coating on the electromagnetic response. By scanning this parameter over the selected interval, we systematically explore how the coating geometry affects both the position and the relative arrangement of the dipolar modes. Figure~\ref{fig:FigTheory}(c,d) shows the normalized wavelengths of the first two dipolar resonances together with the corresponding first antiresonance ($\tilde{\lambda}_{ar}$) as functions of the ratio $t/r$, both without and with a substrate. In all cases, the normalized resonance wavelengths increase with $t/r$, corresponding to a red-shift, or equivalently a decrease in the resonance frequencies. This trend is consistent with the increasing influence of gold. The effect is observed for all modes, but with different strength: the fundamental dipolar mode is the most sensitive to the coating thickness, whereas the higher-order mode varies more weakly. The first antiresonance ($\tilde{\lambda}_{ar}$) exhibits the same overall red-shift with increasing $t/r$, remaining located between the two dipolar resonances. These results show that, once the material contrast is fixed by the choice of Ag and Au, the spectral position of the dipolar resonances can still be efficiently tuned through geometry. The ratio $t/r$ therefore emerges as a key geometrical parameter governing the modal response under realistic material conditions.
While Figure~\ref{fig:FigTheory}(c,d) describe the evolution of the modal positions, they do not directly represent the observable optical response. To establish the connection between the modal analysis and the corresponding far-field spectra, the scattered power was calculated by including the Drude damping rates $\nu_{Au}$ and $\nu_{Ag}$ and by solving the non-homogeneous problem \eqref{eq:DrivenProblem} and then using Eqs. \eqref{eq:Psca}. 
The resulting spectra are shown in Figure~\ref{fig:FigTheory}(e,f) for the same range of the ratio $t/r$, without and with a substrate, respectively. As $t/r$ increases, the low-energy resonance undergoes a pronounced red-shift, while the high-energy resonance shifts much more weakly, consistent with the modal evolution shown in Fig.~\ref{fig:FigTheory}(c,d). By contrast, the antiresonant dip exhibits a limited spectral displacement ($\approx 20$ nm) throughout the investigated range of $t/r$, indicating that the Fano-like interference remains remarkably robust despite the significant evolution of the resonant modes. The introduction of the substrate produces an overall red-shift of the spectra without modifying their qualitative evolution. The characteristic asymmetric Fano-like line shape is therefore preserved, confirming that the antiresonant response originates from the intrinsic modal interference mechanism rather than from the surrounding dielectric environment.

More generally, the present formulation is expressed in terms of the plasma frequencies of the constituent materials. The same framework can therefore be applied to other bimetallic systems by simply replacing the corresponding plasma frequencies and damping rates, without modifying the structure of the model. In this sense, the Ag--Au configuration studied here represents one specific example within a broader class of plasmonic bimetallic nanostructures supporting robust Fano-like optical responses.

\subsection{Experimental Results}

Guided by the theoretical framework introduced above, Au-coated AgNIs were fabricated and characterized after each stage of the fabrication process, namely after Ag deposition (nominal thicknesses of 2 and 5 nm), following thermal annealing, and after deposition of Au overlayers with nominal thicknesses of \(t=3\), 5, 7, and 9 nm. The electrostatic modal analysis identifies the ratio between the Au coating thickness and the NI radius, \(t/r\), as the key geometrical parameter governing the spectral evolution of the hybridized plasmonic modes. In the fabricated system, the nominal Au thickness \(t\) corresponds to the mass-equivalent thickness deposited on a planar reference, as determined by the calibrated deposition rate (0.2 \AA/s) monitored in situ during evaporation. It therefore provides an accurate measure of the total amount of deposited Au. Identifying this nominal thickness with the conformal shell thickness entering the definition of \(t/r\) is, however, an approximation that assumes a nearly uniform Au deposition over the surface of a dense ensemble of non-planar NIs during evaporation, consistent with previous fabrication studies \cite{kvitekPreparationAlloyedCoreshell2020,xuAuAgCoreShell2025}. 

As demonstrated below, this approximation is quantitatively supported by the AFM analysis, confirming that the measured increase in NI size is consistent with the geometrical expectation for a conformal coating. Accordingly, \(t/r\) is treated as an ensemble-averaged effective geometrical parameter throughout the comparison with the electrostatic model. In contrast, the AgNI radius (\(r\)) is established during the solid-state dewetting process and therefore exhibits a statistical distribution. Consequently, increasing the deposited Au thickness shifts the NI population toward larger effective values of \(t/r\), providing an experimental route to investigate the geometrical trends predicted by the modal analysis. 

\begin{figure*}[htbp]
\centering
\includegraphics[width=0.75\textwidth]{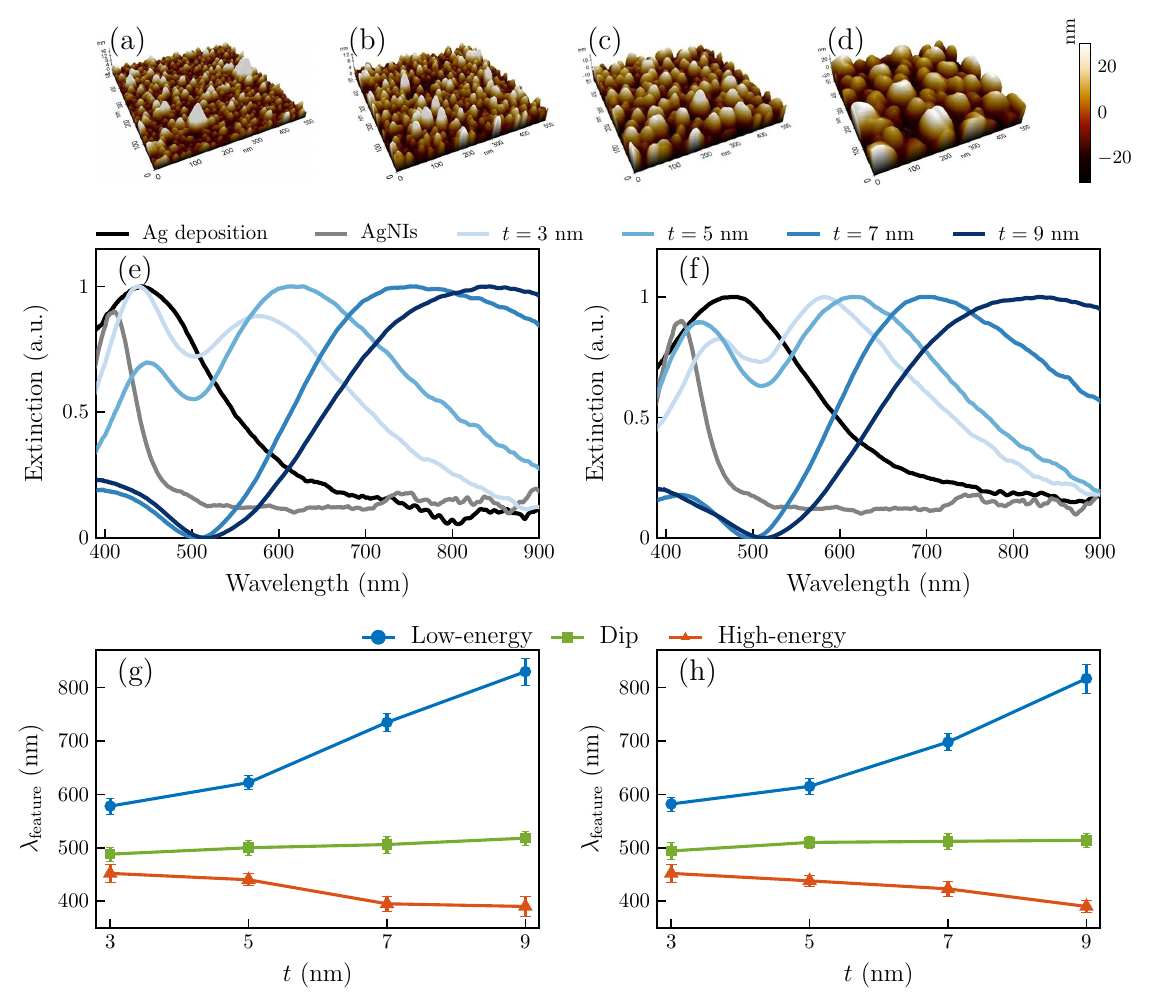}
\caption{
Morphological and optical characterization of AgNIs and Au-coated AgNIs fabricated from two different nominal Ag film thicknesses.
(a--d)~Three-dimensional AFM topographies of the investigated samples; the color bar indicates the height scale in nm.
(a)~AgNIs obtained after thermal dewetting of a 2~nm Ag film,
(b)~the corresponding AgNIs after deposition of a 5~nm nominal Au overlayer,
(c)~AgNIs obtained after dewetting of a 5~nm Ag film, and
(d)~the corresponding AgNIs after deposition of a 5~nm nominal Au overlayer.
(e,f)~Experimental extinction spectra measured after Ag deposition (black), after thermal dewetting into AgNIs (gray), and after deposition of Au overlayers of nominal thickness $t = 3$, $5$, $7$, and $9$~nm, for samples prepared from (e)~2~nm and (f)~5~nm nominal Ag films.
(g,h)~Spectral positions of the low-energy resonance, the Fano-like antiresonant dip, and the high-energy resonance as functions of the deposited Au thickness $t$, for Au-coated AgNIs derived from (g)~AgNIs size $20 \pm 7$~nm and (h)~AgNIs size $34 \pm 10$~nm.
The optical measurements are obtained from a minimum of three independent samples (\(n \geq 3\)). SDs are reported as vertical bars.
}
\label{fig:exp}
\end{figure*}

The optical response was monitored throughout the fabrication steps to investigate how the increase of the effective ratio \(t/r\) influences the extinction spectra. Following Ag deposition, the extinction spectra exhibit a broad and weakly structured profile, characteristic of ultrathin quasi-continuous metallic films \cite{fouadComparativeInsightsStructural2025}. Weak spectral features are observed at approximately \(442 \pm 8\) nm for the 2 nm Ag film and \(477 \pm 13\) nm for the 5 nm Ag film (Figure~\ref{fig:exp}(e,f), black curves), reflecting the onset of plasmonic excitations in partially continuous metallic layers. After thermal annealing, a pronounced spectral reshaping occurs as a consequence of solid-state dewetting, which transforms the continuous films into isolated AgNIs \cite{hruskaSituOpticalMonitoring2025,diernerInfluenceAuAlloying2024}. As a result, well-defined LSPR emerge at approximately \(411 \pm 5\) nm and \(417 \pm 7\) nm for the samples initially deposited with 2 and 5 nm Ag, respectively (Figure~\ref{fig:exp}(e,f), gray curves). The observed blue shift with respect to the as-deposited films is attributed to the transition from a quasi-continuous metallic layer to spatially confined NIs supporting localized plasmonic modes at higher energies \cite{husainPlasmonicNanoislandFilms2026}. 

The morphological evolution underlying the optical response is confirmed by AFM analysis (Figure~\ref{fig:exp}(a--d)). Following thermal annealing, the initially continuous Ag films transform into isolated NIs with average sizes of \(20 \pm 7\) nm and \(34 \pm 10\) nm for the samples prepared from 2 and 5 nm Ag films, respectively. Statistical analysis (Figure~S1) reveals a broad size distribution, reflecting the stochastic nature of the solid-state dewetting process. Following deposition of a nominal Au thickness of \(t=5\) nm, the average NI size increases from \(20 \pm 7\) nm to \(28 \pm 7\) nm (Figure~\ref{fig:exp}(b)) and from \(34 \pm 10\) nm to \(42 \pm 22\) nm (Figure~\ref{fig:exp}(d)), corresponding in both cases to an increase of $\approx 8$ nm. 

Assuming an ideal conformal coating uniformly covering a hemispherical AgNI, the average lateral size is expected to increase by approximately \(2t\). Thus, for \(t=5\,\mathrm{nm}\), an ideal size increase of  $\approx 10$ nm is predicted, whereas the experimentally measured increase of  $\approx 8$ nm for both NI populations corresponds to approximately \(80\%\) of the ideal conformal growth. This slight discrepancy is consistent with moderately sub-conformal lateral coverage and may also be influenced by AFM tip-convolution effects.  

This agreement supports the identification of the nominal deposited thickness with the effective coating thickness adopted in the electrostatic model. In addition, the Au overlayer smooths the NI morphology while broadening the size distribution, consistent with a nearly conformal Au overgrowth. These trends become more pronounced for larger Au thicknesses. Additional AFM and SEM analyses performed on samples coated with \(t=7\) nm are reported in Figure~S2. As the deposited thickness increases, the morphology evolves from isolated NIs toward denser and more interconnected structures. The reduced interparticle separation promotes partial coalescence between neighboring NIs, resulting in broader size distributions and the gradual formation of a quasi-continuous plasmonic network, as confirmed by the SEM observations. Overall, increasing the deposited Au thickness produces two concurrent structural effects. First, it increases the effective ratio \(t/r\), governing the evolution of the hybridized plasmonic modes. Second, it reduces the average interparticle separation, thereby enhancing electromagnetic coupling between neighboring NIs. These morphological changes are expected to influence both the modal properties of individual NIs and their collective optical response. In agreement with this identification, the experimental spectra evolve systematically with increasing \(t\), as shown in Figure~\ref{fig:exp}(e) and Figure~\ref{fig:exp}(f) for the smaller and larger AgNI populations, respectively. 

The optical response progressively broadens and extends from the visible to the near-infrared, while the spectral features become less distinct. This behavior is consistent with the coexistence of two coupled, partially overlapping plasmonic resonances associated with the increasing effective \(t/r\). Part of the observed broadening originates from ensemble effects associated with the larger NI dimensions, the broader size distribution, the enhanced interparticle coupling, and the increased radiative damping accompanying the growth of the Au coating. These effects lie beyond the scope of the present electrostatic model, which treats the NIs as isolated particles. This interpretation is supported by recent studies on disordered plasmonic ensembles, showing that the ensemble linewidth is primarily determined by the distribution of individual-particle resonance frequencies, which broadens with increasing structural heterogeneity and interparticle coupling \cite{shermanDistributionSingleParticleResonances2024}. 

Consequently, the additional inhomogeneous broadening progressively masks the resonance positions at larger \(t\). The measured spectral evolution therefore reflects the combined influence of the increasing effective \(t/r\), modal hybridization, and ensemble broadening associated with the progressive transition from isolated NIs to a strongly coupled plasmonic network. 

To characterize how the spectral positions of the low-energy (superradiant) resonance, the Fano-like antiresonant dip (corresponding to the first antiresonance \(\tilde{\lambda}_{ar}\) of the electrostatic model), and the high-energy (subradiant) resonance depend on \(t\), we track these features in Figure~\ref{fig:exp}(g,h) (see Methods, Sec.~IV.E). As \(t\) increases, the low-energy resonance exhibits a pronounced red-shift, in agreement with the electrostatic model. The high-energy resonance, by contrast, exhibits only a weak apparent blue-shift.

This behavior is not reproduced by the ideal electrostatic model, which neglects spectral broadening arising from material and radiative losses as well as interparticle coupling. In the experimental spectra, the finite resonance linewidths produce a partial spectral overlap between the low-energy and the high energy contributions. For thin coatings, the broad low-energy resonance overlaps the higher-energy feature, shifting its apparent peak position toward longer wavelengths. As t increases, the low-energy resonance red-shifts, reducing the spectral overlap and allowing the apparent high-energy resonance peak to shift slightly toward shorter wavelengths. At larger \(t\), this feature also approaches the short-wavelength limit of the experimental detection range (400 nm), further limiting the accuracy with which its spectral position can be determined. The observed high-energy resonance blue-shift should therefore be interpreted as an apparent shift arising from loss-induced modal overlap and spectral-weight redistribution, rather than as the evolution of an isolated electrostatic eigenmode.

A further feature evident in Figure~\ref{fig:exp}(g,h) is the persistence of the antiresonant minimum separating the low-energy and high-energy resonances across the entire thickness range. Unlike the resonances, whose spectral positions evolve systematically with $t$, the antiresonant dip exhibits a limited spectral displacement throughout the investigated thickness range. In the experiments, the dip shifts by $\approx 23$ nm as the Au coating thickness increases from 3 to 9~nm. This is in agreement with the electrostatic prediction, where both the  antiresonance $\tilde{\lambda}_{ar}$ and the corresponding Drude scattered-power spectra predict a spectral shift of $\approx 20$ nm across the investigated range of $t/r$.

The agreement between the experimental and theoretical spectral evolution of the antiresonant dip provides strong support for the modal-interference interpretation of the Fano-like response. In our system, however, the antiresonant condition [Eq.~\eqref{eq:main_zero}] is an intrinsic consequence of intra-particle modal interference rather than coupling among NIs. Consequently, structural disorder mainly affects the resonance linewidths, whereas the interference-induced minimum remains robust. The resulting asymmetric spectral profile, together with the persistence of the dip and the opposite spectral evolution of the low-energy and high-energy resonances, is characteristic of Fano interference between a broad superradiant mode and a narrower subradiant mode \cite{Fano1961,Lombardi2016}, consistent with previous observations in bimetallic plasmonic nanostructures \cite{maatiExplorationBimetallicAu2022}.

To verify that the observed antiresonant minimum is not an intrinsic feature of the optical response of Au, monometallic AgNIs and AuNIs fabricated using the same protocol were investigated as control samples. Neither sample exhibits the characteristic minimum near 500~nm (Figure~S3), demonstrating that this feature is unique to the Au-coated AgNIs architecture and originates from the coupling between the Ag and Au plasmonic modes rather than from the optical response of either metal alone.

\begin{figure}[htbp!]
\centering
\includegraphics[width=0.75\linewidth]{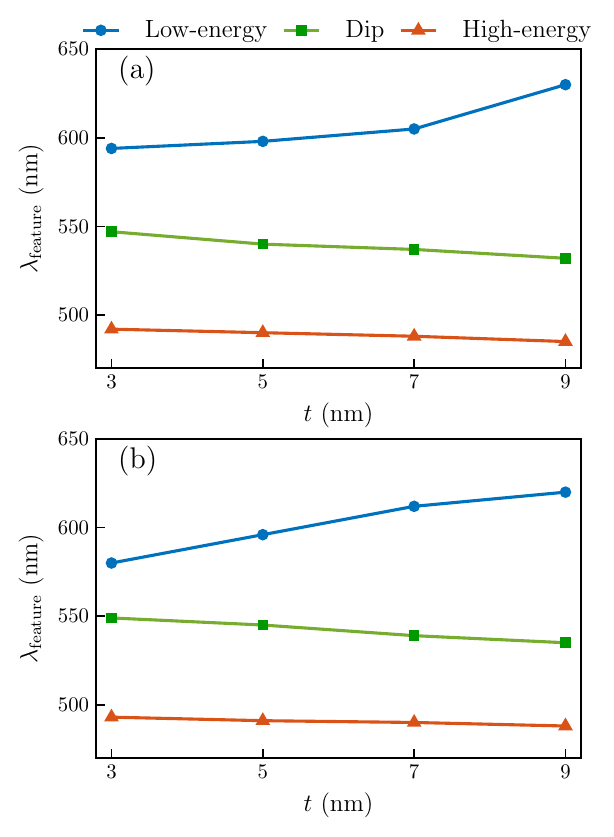}
\caption{Finite element analysis. Extracted spectral positions of the low-energy resonance, the Fano-like antiresonant dip, and the high-energy resonance as a function of the deposited Au thickness \(t\) for hemispherical Au-coated AgNIs with internal radii of (a) \(r=12.5\)  nm and (b) \(r=17.5\) nm.}
\label{fig:comsol1}
\end{figure}

Finally, to account for realistic material dispersion and radiative effects beyond the quasi-electrostatic approximation, we performed full-wave electromagnetic simulations using the finite-element solver COMSOL Multiphysics. Assuming a hemispherical geometry, we considered internal radii of \(r=12.5\) nm  and \(r=17.5\) nm, selected as representative values within the size distributions of the two experimentally observed AgNI populations, and tracked the evolution of the spectral features as a function of the deposited Au thickness (t).

The resulting trends are shown in Figure~\ref{fig:comsol1}. The full-wave calculations reproduce the main experimental trends (Figure~\ref{fig:comsol1} and \ref{fig:exp}(g,h)), including the blue-shift of the high-energy resonance, which is not captured by the quasi-electrostatic model, while confirming weak displacement of the antiresonant dip. The antiresonant dip exhibits a spectral shift of 17 nm, in agreement with both the electrostatic prediction and the experimentally observed shift. Quantitative agreement for the absolute resonance positions and linewidths is not expected because the simulations describe an idealized periodic array of identical Au-coated AgNIs, whereas the experimental spectra arise from an ensemble of NIs and include inhomogeneous broadening associated with the morphological dispersion revealed by AFM, including variations in NI size, shape, interparticle spacing, and local dielectric environment. 

Despite these ensemble effects, the experimental spectra exhibit a systematic and reproducible evolution with increasing Au coating thickness. This confirms that the coating thickness provides a robust parameter for tuning the plasmonic response while preserving the characteristic Fano-like interference.

\section{Conclusions}
\label{sec:Conclusion}

We investigated the origin and tunability of resonant and Fano-like optical features in bimetallic Au-coated AgNIs supported on a substrate and fabricated by solid-state dewetting of thin films. 

We first developed an electrostatic description of the plasmonic modes and resonant frequencies supported by a generic coated NI, assuming a Drude response for both metals. This framework separates the contributions of particle geometry and plasma-frequency mismatch to the resonance condition. By varying either the plasma-frequency contrast or the relative coating thickness, we showed how these parameters govern the hybridized plasmonic modes and the associated antiresonance. By introducing finite Drude damping, the same modal framework was further used to compute the corresponding scattered-power spectra, establishing a direct connection between the electrostatic eigenmodes and the observable Fano-like optical response. 

Then, Au-coated AgNIs were fabricated by thin-film deposition followed by thermally induced solid-state dewetting. Their optical response was characterized by transmission spectroscopy and supported by full-wave numerical simulations. The measured spectra exhibit two resonant features separated by an asymmetric dip. The low-energy peak is mainly associated with the superradiant dipolar mode, whereas the high-energy peak originates from the higher-order subradiant mode. The minimum between them arises from destructive modal interference. Consistent with the electrostatic model, the low-energy resonance undergoes a pronounced red-shift as the Au coating thickness increases. By contrast, the high-energy feature exhibits only a weak blue-shift, which is not reproduced by the lossless electrostatic model but is captured by full-wave finite-element simulations incorporating realistic material dispersion. Most notably, the observed Fano-like antiresonant minimum undergoes a limited spectral displacement of $\approx 23$ nm, in close agreement with the shifts predicted by the electrostatic model $\approx 20$ nm and the full-wave simulations $\approx 17$ nm.

The absence of a comparable antiresonant minimum in the monometallic AgNI and AuNI controls confirms that the Fano-like dip is not an intrinsic feature of either metal, but originates from the coupling and interference between the Ag and Au plasmonic modes in the Au-coated AgNI architecture.

These results establish a comprehensive framework for the rational design of tunable bimetallic plasmonic platforms in which Fano-like interference can be directly accessed through standard linear optical spectroscopy. Despite the morphological disorder intrinsic to solid-state dewetting, the Au-coated AgNI arrays retain a distinct and well-defined antiresonant response, demonstrating the robustness of the underlying modal interference.

The proposed lithography-free fabrication route, based on thin-film deposition and solid-state dewetting \cite{nocerinoNextGenerationPlasmonicPlatforms2025,mirandaMetalEnhancedFluorescenceImmunosensor2020,Bhalla2018,avciTunableFabricationNanoscale2025}, combines wafer-scale scalability with reproducible coating-thickness-controlled spectral tuning. Together with the robustness of the Fano-like response, these characteristics make the platform promising for refractive-index sensing, multiplexed biosensing, and other plasmonic applications.
\newpage

\section{Methods}
\subsection{Boundary-Integral Electrostatic Formulation}

\subsubsection{Derivation of the Electrostatic Eigenvalue Problem}
\label{sec:MethodEigProb}
In this section, we derive the electroquasistatic formulation introduced in Sec.~\ref{sec:EStheory}. The resonance problem is developed following Refs.~\cite{mayergoyz_plasmon_2013,mayergoyz_electrostatic_2005}, whereas the modal-interference analysis follows Ref.~\cite{forestiereTheoryCoupledPlasmon2013}.

We consider the arbitrarily shaped, partially coated NI shown in Fig.~\ref{fig:Sketch_NI}. The geometrical domains, interfaces, and associated unit normals are defined in Sec.~\ref{sec:EStheory} and summarized in Fig.~\ref{fig:Sketch_NI}.

In the quasi-electrostatic limit, the scattering problem can be formulated in terms of the surface-charge densities $\sigma_1$, $\sigma_2$, and $\sigma_3$ supported on the NI--environment interface $S_1$, the NI--coating interface $S_2$, and the coating--environment interface $S_3$, respectively. Following the boundary-integral formulation of Refs.~\cite{mayergoyz_electrostatic_2005,fredkin_resonant_2003, mayergoyz_plasmon_2013}, these surface charges satisfy

\begin{equation}
\epsS (\omega)
\mathcal{A}
\boldsymbol{\sigma}
+
\epsC (\omega) \,
\mathcal{B}
\boldsymbol{\sigma}
+ \mathcal{C}
\boldsymbol{\sigma} = \mathbf{f}_\mathrm{inc}
, 
\label{eq:EigProblem_Sigma}
\end{equation}
where
$\boldsymbol{\sigma}
=
\left[
\sigma_1,
\sigma_2,
\sigma_3
\right]^\intercal$
collects the surface-charge densities on the three interfaces, and
\begin{equation}
\mathbf f_{\mathrm{inc}}(\omega)
=
\begin{bmatrix}
\left[\epsC (\omega)-1\right]\mathbf E_{\mathrm{inc}}\cdot\hat{\mathbf n}_1 \\
\left[\epsC(\omega)-\epsS (\omega)\right]\mathbf E_{\mathrm{inc}}\cdot\hat{\mathbf n}_2 \\
\left[1-\epsS(\omega)\right]\mathbf E_{\mathrm{inc}}\cdot\hat{\mathbf n}_3
\end{bmatrix}.
\label{eq:incident_source}
\end{equation}
is the excitation term.
The geometry-dependent block operators $\mathcal A$, $\mathcal B$, and $\mathcal C$ are
\begin{subequations}
\begin{align}
\mathcal{A} &=  \frac{1}{2}
\begin{bmatrix}
0 & 0 & 0 \\
\mathcal{L}_{21} & \left( \mathcal{L}_{22} + \mathcal{I} \right) &   \mathcal{L}_{23}   \\
-\mathcal{L}_{31}  &   -\mathcal{L}_{32}  &  \left(-\mathcal{L}_{33} + \mathcal{I} \right)
\end{bmatrix},
\label{eq:OperatorA} \\
%%%%%%%%%%%%%%%%%%%%
\mathcal{B} &=   \frac{1}{2}
\begin{bmatrix}
 \left( \mathcal{I} - \mathcal{L}_{11} \right)  & - \mathcal{L}_{12}  &  - \mathcal{L}_{13}   \\
-\mathcal{L}_{21}  &  \left( -\mathcal{L}_{22} + \mathcal{I} \right)   &  -\mathcal{L}_{23}   \\
0 & 0 & 0  
\end{bmatrix}, 
\label{eq:OperatorB}\\
\mathcal{C} &= \frac{1}{2}
\begin{bmatrix}
\left(  \mathcal{I} + \mathcal{L}_{11} \right) &   \mathcal{L}_{12} &   \mathcal{L}_{13}   \\
0 & 0 & 0 \\
\mathcal{L}_{31}  &   \mathcal{L}_{32}  &  \left(\mathcal{L}_{33} +  \mathcal{I}  \right)
\end{bmatrix}. 
\label{eq:OperatorC}
\end{align}
\end{subequations}
Here, $\mathcal I$ denotes the identity operator, while $ \mathcal{L}_{uv}: \mathbb{L}^2 \left( S_v \right) \rightarrow \mathbb{L}^2 \left(
S_u	\right)$   is defined $\forall u,v \in \left\{1,2,3\right\}$ as 
\begin{equation}
 \mathcal{L}_{uv} \left\{  \sigma \right\}\left(Q \right)  =   \frac{1}{2\pi} \oint_{S_v} \sigma \left( M \right) \frac{{\bf r}_{MQ}
\cdot {\bf n}_Q
}{{r}_{MQ}^3} dS_M, \quad Q \in S_u.
 \label{eq:Operator_L}
\end{equation}
%% WC 62
Here, \(\mathbf r_{MQ}=\mathbf r_Q-\mathbf r_M\), \(r_{MQ}=|\mathbf r_{MQ}|\), and \(\mathbf n_Q\) is the unit normal at \(Q\). By direct inspection, one finds that $\mathcal{A} +     \mathcal{B} +     \mathcal{C} = \mathcal{I}$.

We now assume that the NI and coating obey the Drude dispersion relations introduced in Eq.~\eqref{eq:DrudeDispersion}. In the lossless limit, obtained by setting $\nu_{\mathrm{I}}=\nu_{\mathrm{C}}=0$,  Eq.~\eqref{eq:EigProblem_Sigma} reduces to

\begin{equation}
  \left[ \omegapS^2  \mathcal{A} +     \omegapC^2  \mathcal{B} \right]  \boldsymbol{\sigma} = \omega^2  \boldsymbol{\sigma} - \mathbf g_{\mathrm{inc}}
  \label{eq:EigProblem0}
\end{equation}

\begin{equation}
\mathbf g_{\mathrm{inc}}
=
\begin{bmatrix}
-{\omega}_{p,\mathrm I}^2\,\mathbf E_{\mathrm{inc}}\cdot\hat{\mathbf n}_1 \\
\left(\omegapS^2-{\omega}_{p,\mathrm I}^2\right)\mathbf E_{\mathrm{inc}}\cdot\hat{\mathbf n}_2 \\
{\omega}_{p,\mathrm C}^2\,\mathbf E_{\mathrm{inc}}\cdot\hat{\mathbf n}_3
\end{bmatrix}.
\label{eq:c_inc}
\end{equation}
 We  now introduce the normalized plasma frequency mismatch $\delta$,  defined by Eq. \eqref{eq:main_delta}. The physical meaning of $\delta$ is discussed in Sec.~\ref{sec:EStheory}. We normalize all angular frequencies to the root-mean-square plasma frequency defined in Eq.~\eqref{eq:root_mean_square}, namely: 
\begin{equation}
\tilde{\omega} =  \frac{\omega}{\omegarms}, \quad
\tilde{\omega}_{p,\mathrm C}^2
=
\frac{\omegapS^2}{\omegarms^2}
=
1+\delta,
\quad
\tilde{\omega}_{p,\mathrm I}^2
=
\frac{\omega_{p,\mathrm I}^2}{\omegarms^2}
=
1-\delta.
\end{equation}
 
 Dividing Eq.~\eqref{eq:EigProblem0} by $\omega_{\mathrm{rms}}^2$ gives
\begin{equation}
  \left[\mathcal{E} +     \delta  \, \mathcal{F} \right]  \boldsymbol{\sigma} = \tilde{\omega}^2  \boldsymbol{\sigma} - \tilde{\mathbf g}_{\mathrm{inc}}
  \label{eq:NonHOMProblem}
\end{equation}
where $\mathcal{E} =   \left( \mathcal{A} + \mathcal{B} \right)$ and $\mathcal{F} =  \left( \mathcal{A} - \mathcal{B} \right)$, i.e.
%%%%%%%%%%%%%
% Integral Operator E
%%%%%%%%%%%%%
\begin{subequations}
\begin{align}
\mathcal{E} &=    \frac{1}{2}
\begin{bmatrix}
 \left( \mathcal{I} - \mathcal{L}_{11} \right)  & - \mathcal{L}_{12}  &  - \mathcal{L}_{13}   \\
 0 & 2\mathcal{I}  &  0  \\
-\mathcal{L}_{31}  &   -\mathcal{L}_{32}  &  \left(-\mathcal{L}_{33} + \mathcal{I} \right)
\end{bmatrix}, 
\label{eq:IntegralOperatorE} \\\
\mathcal{F} &=   \frac{1}{2}
\begin{bmatrix}
 \left( \mathcal{L}_{11}   - \mathcal{I} \right)  & + \mathcal{L}_{12}  &  + \mathcal{L}_{13}   \\
2\mathcal{L}_{21} & 2 \mathcal{L}_{22}  &   2\mathcal{L}_{23}   \\
-\mathcal{L}_{31}  &   -\mathcal{L}_{32}  &  \left(-\mathcal{L}_{33} + \mathcal{I} \right)
\end{bmatrix}, 
\label{eq:IntegralOperatorF}
\end{align}
\end{subequations}
and the normalized excitation is \(\tilde{\mathbf g}_{\mathrm{inc}}=\mathbf g_{\mathrm{inc}}/\omegarms^2\), i.e.
\begin{equation}
\tilde{\mathbf g}_{\mathrm{inc}}
=
\begin{bmatrix}
(\delta -1 )\,\mathbf E_{\mathrm{inc}}\cdot\hat{\mathbf n}_1 \\
2 \delta \, \mathbf E_{\mathrm{inc}}\cdot\hat{\mathbf n}_2 \\
(\delta + 1) \,\mathbf E_{\mathrm{inc}}\cdot\hat{\mathbf n}_3
\end{bmatrix}.
\label{eq:c_inc}
\end{equation}

The electrostatic resonances are obtained from the source-free problem by setting the excitation vector to zero:
\begin{equation}
\left(\mathcal E+\delta\mathcal F\right)
\boldsymbol{\sigma}^{(k)}
=
\tilde{\omega}_k^2
\boldsymbol{\sigma}^{(k)}.
\label{eq:eigenproblem}
\end{equation}
The eigenvalues $\widetilde{\omega}_k$ are the normalized resonance frequencies, while
$
\boldsymbol{\sigma}^{(k)}
=
\begin{bmatrix}
\sigma_1^{(k)} &
\sigma_2^{(k)} &
\sigma_3^{(k)}
\end{bmatrix}^{\mathsf T}
$
collects the corresponding surface-charge distributions on $S_1$, $S_2$, and $S_3$. Equation~\eqref{eq:eigenproblem} separates the geometrical contribution, contained in $\mathcal E$ and $\mathcal F$, from the material asymmetry controlled by $\delta$. In the material-symmetric limit $\delta=0$, the spectrum is governed solely by $\mathcal E$.

Because Eq.~\eqref{eq:eigenproblem} is purely electrostatic, its eigenvalues are scale invariant: a uniform scaling of all geometrical dimensions leaves \(\tilde{\omega}_k\) unchanged. Therefore, for a fixed shape, the normalized spectrum depends on dimensionless geometrical ratios and on the mismatch parameter \(\delta\), but not on the absolute particle size. The dimensional resonance frequencies are then recovered through \(\omega_k=\tilde{\omega}_k\omegarms\).

\subsubsection{Inclusion of a Dielectric Substrate}
\label{sec:substrate}
For a NI supported by a dielectric substrate, the free-space kernel in Eq.~\eqref{eq:Operator_L} is replaced by the corresponding half-space Green function:
\begin{equation}
\mathcal L_{uv}\{\sigma\}(Q)=\frac{1}{2\pi}\oint_{S_v}
\sigma(M)\,\mathbf n_Q\cdot\nabla_Q G(Q,M)\,dS_M ,
\end{equation}
with
\begin{equation}
G(Q,M)=\frac{1}{r_{MQ}}
-\frac{\varepsilon_{r,\mathrm{sub}}-1}
{\varepsilon_{r,\mathrm{sub}}+1}\,
\frac{1}{r_{M'Q}} .
\end{equation}
Here, \(\varepsilon_{r,\mathrm{sub}}\) is the relative permittivity of the substrate,  and \(M'\) is the image point of \(M\) with respect to the substrate plane.

%Finally, we stress that the Drude reduction above is intended to isolate the role of plasma-frequency mismatch in the electrostatic eigenvalue problem. It does not capture the interband contribution of Au in the visible range; that physics is incorporated later through full-wave calculations using realistic optical data. 

\subsubsection{Modal Excitation and Dipolar Interference}
\label{sec:coupling}
To analyze modal interference, we return to the forced problem in Eq.~\eqref{eq:NonHOMProblem}, and expand its solution in the basis of
right charge-density eigenmodes $\left\{ \boldsymbol{\sigma}^{(h)} \right\}$. Since the operator \(\mathcal E+\delta\mathcal F\) is generally non-Hermitian, the
corresponding left eigenmodes  $\left\{ \boldsymbol{\tau}^{(k)} \right\}$ must also be introduced. They are defined by through the adjoint eigenvalue problem \cite{mayergoyz_electrostatic_2005}
\begin{equation}
\left(\mathcal E+\delta\mathcal F\right)^\dagger
\boldsymbol{\tau}^{(k)}
=
\tilde{\omega}_k^2
\boldsymbol{\tau}^{(k)}.
\label{eq:adjoint_problem}
\end{equation}
For the electrostatic eigenvalue problem considered here, the eigenvalues \(\tilde{\omega}_k^2\) are real, so that the direct and adjoint problems share the same spectrum. Analogously to the right eigenmodes, each left eigenmode is represented as
$
\boldsymbol{\tau}^{(k)} = 
\left[
\tau^{(k)},
\tau^{(k)},
\tau^{(k)}   
\right]^\intercal
$ 
where the three components are defined on the interfaces \(S_1\), \(S_2\), and \(S_3\), respectively. It is important to note that the adjoint problem admit the same eigenvalues while the eigenvectors \(\boldsymbol{\tau}^{(k)}\) and \(\boldsymbol{\sigma}^{(k)}\) associated to different eigenvalues are orthogonal, namely 

\begin{equation}
    \left\langle \boldsymbol{\tau}^{(k)}, \boldsymbol{\sigma}^{(h)} \right\rangle = 0 \qquad \text{for} \quad \tilde{\omega}_h \ne \tilde{\omega}_k,
\end{equation}
where the inner product is taken over the three interfaces,
\begin{equation}
\left\langle \mathbf u,\mathbf v\right\rangle
=
\sum_{i=1}^{3}
\int_{S_i}
u_i^*(M)v_i(M)\,dS_M .
\end{equation}

Expanding the induced charge density solution of the non homogeneous problem Eq.~\eqref{eq:NonHOMProblem} in the eigenmode basis,
\begin{equation}
\boldsymbol{\sigma}
=
\sum_{k=1}^{\infty}
a_k\boldsymbol{\sigma}^{(k)},
\label{eq:modal_expansion}
\end{equation}
and projecting Eq.~\eqref{eq:NonHOMProblem} onto the adjoint modes  \(\boldsymbol{\tau}^{(k)}\) gives
\begin{equation}
\boldsymbol{\sigma}
=
\sum_{k=1}^{\infty}
\frac{\alpha_k}
{\tilde{\omega}_k^2-\tilde{\omega}^2}
\boldsymbol{\sigma}^{(k)},
\label{eq:sigma_modal_solution}
\end{equation}
where
\begin{equation}
\alpha_k
=
-
\frac{
\left\langle
\boldsymbol{\tau}^{(k)},
\tilde{\mathbf g}_{\mathrm{inc}}
\right\rangle
}{
\left\langle
\boldsymbol{\tau}^{(k)},
\boldsymbol{\sigma}^{(k)}
\right\rangle
}.
\label{eq:alpha_k}
\end{equation}
is the excitation coefficient of the \(k\)-th mode.

In the dipolar approximation, the induced electric dipole moment of the coated particle is
\begin{equation}
\mathbf p
=
\sum_{i=1}^{3}
\int_{S_i}
\mathbf r\,\sigma_i(\mathbf r)\,dS .
\label{eq:dipole_moment}
\end{equation}
The dipole moment associated with the \(k\)-th eigenmode is
\begin{equation}
\mathbf p^{(k)}
=
\sum_{i=1}^{3}
\int_{S_i}
\mathbf r\,\sigma_i^{(k)}(\mathbf r)\,dS .
\label{eq:modal_dipole}
\end{equation}
Substituting the modal expansion
\eqref{eq:sigma_modal_solution} into Eq.~\eqref{eq:dipole_moment} and using
Eq.~\eqref{eq:modal_dipole}, the \(t\)-component of the induced dipole moment, i.e. $p_t
= \mathbf{p} \cdot \hat{\mathbf{t}}$, becomes
\begin{equation}
p_t(\tilde{\omega})
=
\sum_{k=1}^{\infty}
\frac{\alpha_k }
{\tilde{\omega}_k^2-\tilde{\omega}^2} p_t^{(k)}, \qquad \text{with} \quad t=x,y,z.
\label{eq:dipole_modal_expansion}
\end{equation}
Assuming $t = x$ we obtain Eq. \eqref{eq:main_modal_dipole} of the main text.

\subsection{Fabrication of Au-coated AgNIs}
Au-coated AgNIs were fabricated by adapting previously reported protocols \cite{Bhalla2018,mirandaMetalEnhancedFluorescenceImmunosensor2020,nocerinoNextGenerationPlasmonicPlatforms2025, ametrano2024structural} in a class 1000 cleanroom environment. Glass coverslips (15 mm $\times$ 15 mm) were cleaned by sequential ultrasonication in acetone and isopropanol for 2 min each, followed by drying under a nitrogen stream. A polydimethylsiloxane (PDMS) shadow mask with circular openings (8 mm in diameter) was placed on the substrates to define the deposition area.

The masked substrates were introduced into the high-vacuum chamber of a thermal evaporator (LEYBOLD HERAEUS L 560). After reaching a base pressure of $\sim10^{-6}$~mbar, Ag thin films with nominal thicknesses of 2 and 5~nm were deposited at a rate of 0.2~\AA~s$^{-1}$. The samples were then annealed at 560~\textdegree C for 3.5~h to induce solid-state dewetting of the Ag film, resulting in the formation of isolated AgNIs.

An Au overlayer of thickness $t$ was subsequently deposited onto the AgNIs by thermal evaporation at room temperature. The Au coating thickness was tuned by depositing nominal Au films of 3, 5, 7, and 9~nm. No post-deposition annealing was performed to prevent Ag--Au interdiffusion and alloy formation, thereby preserving the intended coating architecture.

\subsection{Optical Characterization}

The optical extinction response of the fabricated Au-coated AgNIs was characterized by normal-incidence far-field transmission spectroscopy using a custom-built setup \cite{nocerinoNextGenerationPlasmonicPlatforms2025}. Broadband illumination from a halogen lamp was coupled into a Thorlabs optical fiber and collimated onto the sample, providing uniform illumination over the nanostructured area with a spot diameter of approximately 2~mm. The transmitted light was collected by a second Thorlabs optical fiber and analyzed with a spectrometer (Filmetrics F20), covering the 400--900~nm wavelength range with a spectral resolution of approximately 0.5~nm.

The samples were mounted in a dedicated holder to ensure reproducible alignment. The illumination spot size and the separation between the illumination and collection fibers were kept constant throughout all measurements. Reference spectra acquired from a bare glass substrate were used to determine the optical extinction, defined as
\[
E(\lambda)=-\log_{10}[T(\lambda)],
\]
where \(T(\lambda)\) is the transmittance normalized to the bare-glass reference.

For each experimental condition, measurements were performed on at least three independently fabricated samples (\(n \geq 3\)). Error bars represent the standard deviation of the independently measured spectra. The spectral positions of the low-energy resonance, the antiresonant dip, and the high-energy resonance were extracted as local extrema of the raw (unsmoothed) extinction spectra using OriginPro. For the largest investigated Au coating thickness (\(t = 9\)~nm), the low-energy resonance becomes significantly broadened and no longer exhibits a well-defined maximum, consistent with the evolution of the NI ensemble toward a denser, partially coalesced morphology. Consequently, the extracted peak position carries a larger uncertainty, as reflected by the corresponding error bar in Fig.~\ref{fig:exp}(g,h).

\subsection{AFM and SEM Morphological Characterization}

The morphology of the fabricated bimetallic nanostructures was primarily characterized by Atomic Force Microscopy (AFM), which was employed to investigate the surface topography and to perform statistical grain-size analysis of the nanostructures. Scanning Electron Microscopy (SEM) was additionally used as a complementary technique to support the AFM observations and to further assess the morphology of selected samples at different stages of the fabrication process. SEM images were acquired using a 25 pA electron beam operated at an accelerating voltage of 10 kV and at different magnifications. The samples were mounted on aluminum stubs using conductive carbon tape.

\subsection{Full-Wave Electromagnetic Simulations}

Full-wave electromagnetic simulations were performed using COMSOL Multiphysics to model the optical response of a periodic array of Au-coated AgNIs. The computational domain consisted of a square unit cell with lateral period \(d_x=d_y=150~\mathrm{nm}\), containing a single hemispherical Au-coated AgNI supported on a glass substrate and surrounded by air. Floquet periodic boundary conditions were applied on the lateral boundaries to represent the periodic repetition of the unit cell, while plane-wave excitation was introduced through port boundary conditions placed above and below the structure, which also terminated the semi-infinite air and substrate domains.
The AgNI was described using the optical constants available in the COMSOL material library (Raki\'{c} \textit{et al.}, 1998, based on the Brendel--Bormann model), while the conformal Au coating was modeled using the corresponding optical constants from the same library. The glass substrate was treated as a homogeneous dielectric medium with refractive index \(n=1.45\). The computational domain was discretized using COMSOL's physics-controlled ``Finer'' mesh.
The optical response was calculated by performing parametric sweeps over the excitation wavelength, the NI radius, and the Au coating thickness. Simulations were carried out over the wavelength range 400--800~nm, covering the spectral region relevant to the plasmonic response of Ag and Au. The internal radii were set to \(r = 12.5\) nm and \(r = 17.5\) nm, matching the mean radii of the two AFM-measured NI populations, while remaining compatible with the adopted unit-cell period.
 
% Clean ACS Photonics-style bibliography generated from the supplied .bbl file.
% Insert this block immediately before \end{document}.
% Do not also use \bibliography{...} or \printbibliography in the same document.

\begin{thebibliography}{63}

\bibitem{haMulticomponentPlasmonicNanoparticles2019}
Ha,~M.; Kim,~J.-H.; You,~M.; Li,~Q.; Fan,~C.; Nam,~J.-M. Multicomponent
  plasmonic nanoparticles: from heterostructured nanoparticles to colloidal
  composite nanostructures. \emph{Chem. Rev.} \textbf{2019}, \emph{119},
  12208--12278.

\bibitem{sytwu_bimetallic_2019}
Sytwu,~K.; Vadai,~M.; Dionne,~J.~A. Bimetallic nanostructures: combining
  plasmonic and catalytic metals for photocatalysis. \emph{Adv. Phys. X} \textbf{2019}, \emph{4}, 1619480.

\bibitem{chakrabortyEvolutionColloidalPlasmonic2024}
Chakraborty,~S.; Das,~S.; Dash,~G.; Viswanatha,~R. Evolution of colloidal
  plasmonic heterostructures from traditional semiconductor nanocrystals to
  lead halide perovskites: a review. \emph{ACS Appl. Nano Mater.}
  \textbf{2024}, \emph{7}, 2494--2514.

\bibitem{sahaElectronEnergyLoss2024}
Saha,~S.~K.; Mondal,~P.; Channagiri,~S.~A.; Mahadevu,~R.; Vasudeva,~N.;
  Bellare,~P.; Ravishankar,~N.; Pandey,~A. Electron energy loss spectroscopic
  investigation of {Mie} resonances in bimetallic nanostructures.
  \emph{Chem. Phys. Impact} \textbf{2024}, \emph{9}, 100677.

\bibitem{prodanHybridizationModelPlasmon2003}
Prodan,~E.; Radloff,~C.; Halas,~N.~J.; Nordlander,~P. A hybridization model for
  the plasmon response of complex nanostructures. \emph{Science} \textbf{2003},
  \emph{302}, 419--422.

\bibitem{mamoGeometricDielectricModulation2025}
Mamo,~S.~G. Geometric and dielectric modulation of nonlinear optical properties
  in {ZnTe}@ {Ag} core--shell nanostructures: a comparative study of spherical
  and cylindrical inclusions. \emph{Eur. Phys. J. D}
  \textbf{2025}, \emph{79}, 107.

\bibitem{heAgAuBimetallic2025}
He,~S.; Tang,~Z.; Huo,~T.; Wu,~D.; Tang,~J.~H. Ag/{Au} bimetallic core--shell
  nanostructures: a review of synthesis and applications. \emph{J. Manuf. Mater. Process.} \textbf{2025}, \emph{9}, 131.

\bibitem{awiazRecentAdvancesAu2023}
Awiaz,~G.; Lin,~J.; Wu,~A. Recent advances of {Au}@ {Ag} core--shell
  {SERS}-based biosensors. 2023; p 20220072.

\bibitem{tuff_ion_2023}
Tuff,~W.~J.; Hughes,~R.~A.; Golze,~S.~D.; Neretina,~S. Ion {Beam} {Milling} as
  a {Symmetry}-{Breaking} {Control} in the {Synthesis} of {Periodic} {Arrays}
  of {Identically} {Aligned} {Bimetallic} {Janus} {Nanocrystals}. \emph{ACS Nano} \textbf{2023}, \emph{17}, 4050--4061.

\bibitem{kangStabilizationSilverGold2018}
Kang,~H.; Buchman,~J.~T.; Rodriguez,~R.~S.; Ring,~H.~L.; He,~J.; Bantz,~K.~C.;
  Haynes,~C.~L. Stabilization of silver and gold nanoparticles: preservation
  and improvement of plasmonic functionalities. \emph{Chem. Rev.}
  \textbf{2018}, \emph{119}, 664--699.

\bibitem{lozaSynthesisStructureProperties2020}
Loza,~K.; Heggen,~M.; Epple,~M. Synthesis, structure, properties, and
  applications of bimetallic nanoparticles of noble metals. \emph{Adv. Funct. Mater.} \textbf{2020}, \emph{30}, 1909260.

\bibitem{ghoshchaudhuriCoreShellNanoparticles2012}
Ghosh~Chaudhuri,~R.; Paria,~S. Core/shell nanoparticles: classes, properties,
  synthesis mechanisms, characterization, and applications. \emph{Chem. Rev.} \textbf{2012}, \emph{112}, 2373--2433.

\bibitem{bonviciniFormationRemovalAlloyed2022}
Bonvicini,~S.~N.; Shi,~Y. Formation and removal of alloyed bimetallic
  {Au}--{Ag} nanoparticles from silicon substrates for tunable surface plasmon
  resonance. \emph{ACS Appl. Nano Mater.} \textbf{2022}, \emph{5},
  14850--14861.

\bibitem{ametrano2024structural}
Ametrano,~A.; Miranda,~B.; Moretta,~R.; Dardano,~P.; De~Stefano,~L.;
  Oreste,~U.; Coscia,~M.~R. A structural peculiarity of Antarctic fish IgM
  drives the generation of an engineered mAb by CRISPR/Cas9. \emph{Front. Bioeng. Biotechnol.} \textbf{2024}, \emph{12}, 1315633.

\bibitem{luSwitchingPlasmonicFano2019}
Lu,~W.; Cui,~X.; Chow,~T.~H.; Shao,~L.; Wang,~H.; Chen,~H.; Wang,~J. Switching
  plasmonic {Fano} resonance in gold nanosphere--nanoplate heterodimers.
  \emph{Nanoscale} \textbf{2019}, \emph{11}, 9641--9653.

\bibitem{zorattiSimpleOneStepSynthesis2025}
Zoratti,~M.; Krepper,~G.; Hernandez,~S.~A.; Reinoso,~D.~M. Simple {One}-{Step}
  {Synthesis} of {Bimetallic} {Au}--{Ag} {Nanoparticles}: {Integrated}
  {Nanoscience} {Laboratory} {Project}. \emph{J. Chem. Educ.}
  \textbf{2025}, \emph{102}, 1598--1604.

\bibitem{weller_gap-dependent_2016}
Weller,~L.; Thacker,~V.~V.; Herrmann,~L.~O.; Hemmig,~E.~A.; Lombardi,~A.;
  Keyser,~U.~F.; Baumberg,~J.~J. Gap-{Dependent} {Coupling} of {Ag}--{Au}
  {Nanoparticle} {Heterodimers} {Using} {DNA} {Origami}-{Based}
  {Self}-{Assembly}. \emph{ACS Photonics} \textbf{2016}, \emph{3},
  1589--1595.

\bibitem{aldufeeryDisentanglingBrightDark2026}
Aldufeery,~E.~A. Disentangling bright and dark plasmonic modes in hexagonal
  nanoprism oligomers. \emph{Photonics Nanostruct. Fundam. Appl.} \textbf{2026}, \emph{71}, 101545.

\bibitem{srnova-sloufova_coreshell_2000}
Srnová-Šloufová,~I.; Lednický,~F.; Gemperle,~A.; Gemperlová,~J.
  Core-{Shell} ({Ag}){Au} {Bimetallic} {Nanoparticles}: {Analysis} of
  {Transmission} {Electron} {Microscopy} {Images}. \emph{Langmuir}
  \textbf{2000}, \emph{16}, 9928--9935.

\bibitem{rodriguez-gonzalez_multishell_2005}
Rodríguez-González,~B.; Burrows,~A.; Watanabe,~M.; Kiely,~C.~J.;
  Liz~Marzán,~L.~M. Multishell bimetallic {AuAg} nanoparticles: synthesis,
  structure and optical properties. \emph{J. Mater. Chem.}
  \textbf{2005}, \emph{15}, 1755--1759.

\bibitem{zhang_surface_2015}
Zhang,~C.; Chen,~B.-Q.; Li,~Z.-Y.; Xia,~Y.; Chen,~Y.-G. Surface {Plasmon}
  {Resonance} in {Bimetallic} {Core}--{Shell} {Nanoparticles}. \emph{J. Phys. Chem. C} \textbf{2015}, \emph{119},
  16836--16845.

\bibitem{hangPlasmonicSilverGold2024}
Hang,~Y.; Wang,~A.; Wu,~N. Plasmonic silver and gold nanoparticles: shape-and
  structure-modulated plasmonic functionality for point-of-caring sensing,
  bio-imaging and medical therapy. \emph{Chem. Soc. Rev.}
  \textbf{2024}, \emph{53}, 2932--2971.

\bibitem{muhammadDualOptimizationShell2025}
Muhammad,~S.; Shujah,~S.; Gao,~S.; Singh,~V.; Rong,~H.; Che,~J.; Zhang,~J. Dual
  {Optimization} of {Shell} and {Interparticle} {Gaps} in {Plasmonic} {Au}@
  {Ag} {Nanocubes} {Assembly} for {Hot} {Spot}-{Driven} {SERS} {Performance}.
  \emph{Inorg. Chem.} \textbf{2025}, \emph{64}, 24529--24538.

\bibitem{liHighlyHomogeneousBimetallic2023}
Li,~S.; Chen,~J.; Xu,~W.; Sun,~B.; Wu,~J.; Chen,~Q.; Liang,~P. Highly
  homogeneous bimetallic core--shell {Au}@ {Ag} nanoparticles with embedded
  internal standard fabrication using a microreactor for reliable quantitative
  {SERS} detection. \emph{Mater. Chem. Front.} \textbf{2023},
  \emph{7}, 1100--1109.

\bibitem{lukyanchukFanoResonancePlasmonic2010}
Luk'Yanchuk,~B.; Zheludev,~N.~I.; Maier,~S.~A.; Halas,~N.~J.; Nordlander,~P.;
  Giessen,~H.; Chong,~C.~T. The {Fano} resonance in plasmonic nanostructures
  and metamaterials. \emph{Nat. Mater.} \textbf{2010}, \emph{9},
  707--715.

\bibitem{Miroshnichenko2010}
Miroshnichenko,~A.~E.; Flach,~S.; Kivshar,~Y.~S. Fano resonances in nanoscale
  structures. \emph{Rev. Mod. Phys.} \textbf{2010}, \emph{82},
  2257--2298.

\bibitem{zebFanoResonanceStrongcoupling2022}
Zeb,~M.~A. Fano resonance in the strong-coupling regime. \emph{Phys. Rev. B} \textbf{2022}, \emph{106}, 155134.

\bibitem{wangTunableFanoResonance2018}
Wang,~M.; Krasnok,~A.; Zhang,~T.; Scarabelli,~L.; Liu,~H.; Wu,~Z.;
  Liz-Marzán,~L.~M.; Terrones,~M.; Alù,~A.; Zheng,~Y. Tunable fano
  resonance and plasmon--exciton coupling in single au nanotriangles on
  monolayer {WS2} at room temperature. \emph{Adv. Mater.} \textbf{2018},
  \emph{30}, 1705779.

\bibitem{forestiereTheoryCoupledPlasmon2013}
Forestiere,~C.; Dal~Negro,~L.; Miano,~G. Theory of coupled plasmon modes and
  {Fano}-like resonances in subwavelength metal structures. \emph{Phys. Rev. B} \textbf{2013}, \emph{88},
  155411.

\bibitem{bachelierFanoProfilesInduced2008}
Bachelier,~G.; Russier-Antoine,~I.; Benichou,~E.; Jonin,~C.; Del~Fatti,~N.;
  Vallée,~F.; Brevet,~P.-F. Fano {Profiles} {Induced} by {Near}-{Field}
  {Coupling} in {Heterogeneous} {Dimers} of {Gold} and {Silver}
  {Nanoparticles}. \emph{Phys. Rev. Lett.} \textbf{2008}, \emph{101},
  197401.

\bibitem{Lombardi2016}
Lombardi,~A.; Grzelczak,~M.~P.; Pertreux,~E.; Crut,~A.; Maioli,~P.;
  Pastoriza-Santos,~I.; Liz-Marz{\'a}n,~L.~M.; Vall{\'e}e,~F.; Del~Fatti,~N.
  Fano Interference in the Optical Absorption of an Individual Gold--Silver
  Nanodimer. \emph{Nano Lett.} \textbf{2016}, \emph{16}, 6311--6316.

\bibitem{zhuMultipleFanoResonances2026}
Zhu,~X.; Mi,~J.; Hu,~W.; Xu,~S.; He,~H.; Hu,~J. Multiple fano resonances in
  all-dielectric metasurface for bidirectional optical switching and
  high-performance multichannel sensing. \emph{Opt. Commun.}
  \textbf{2026}, 133327.

\bibitem{alvesEnhancedSurfaceFields2024}
Alves,~R.~S.; Mazali,~I.~O.; dos Santos,~D.~P. Enhanced {Surface} {Fields}
  {Driven} by {Fano} {Resonances} in {Silver} {Nanocube} {Dimers} for
  {Efficient} {Hot} {Electron} {Generation}. \emph{J. Phys. Chem. C} \textbf{2024}, \emph{128}, 9182--9192.

\bibitem{barelliColorRoutingCrosspolarized2020}
Barelli,~M.; Mazzanti,~A.; Giordano,~M.~C.; Della~Valle,~G.; Buatier~de
  Mongeot,~F. Color routing via cross-polarized detuned plasmonic nanoantennas
  in large-area metasurfaces. \emph{Nano Lett.} \textbf{2020}, \emph{20},
  4121--4128.

\bibitem{leeControlledAssemblyPlasmonic2021}
Lee,~S.; Sim,~K.; Moon,~S.~Y.; Choi,~J.; Jeon,~Y.; Nam,~J.; Park,~S. Controlled
  assembly of plasmonic nanoparticles: from static to dynamic nanostructures.
  \emph{Adv. Mater.} \textbf{2021}, \emph{33}, 2007668.

\bibitem{changPlasmonicFanoSwitch2012}
Chang,~W.-S.; Lassiter,~J.~B.; Swanglap,~P.; Sobhani,~H.; Khatua,~S.;
  Nordlander,~P.; Halas,~N.~J.; Link,~S. A plasmonic {Fano} switch. \emph{Nano Lett.} \textbf{2012}, \emph{12}, 4977--4982.

\bibitem{attiaouiPolarizationTunedFanoResonances2023}
Attiaoui,~A.; Daligou,~G.; Assali,~S.; Skibitzki,~O.; Schroeder,~T.;
  Moutanabbir,~O. Polarization-{Tuned} {Fano} {Resonances} in
  {All}-{Dielectric} {Short}-{Wave} {Infrared} {Metasurface}.
  \emph{Adv. Mater.} \textbf{2023}, \emph{35}, 2300595.

\bibitem{liPhaseControlledFanoResonances2026}
Li,~M.; Zhong,~Z.; Bai,~P.; Peng,~S. Phase-{Controlled} {Fano} {Resonances} in
  {Hybrid} {Metasurfaces}. \emph{ACS Nano} \textbf{2026}, \emph{20},
  8406--8414.

\bibitem{flauraud_mode_2017}
Flauraud,~V.; Bernasconi,~G.~D.; Butet,~J.; Alexander,~D. T.~L.; Martin,~O.
  J.~F.; Brugger,~J. Mode {Coupling} in {Plasmonic} {Heterodimers} {Probed}
  with {Electron} {Energy} {Loss} {Spectroscopy}. \emph{ACS Nano}
  \textbf{2017}, \emph{11}, 3485--3495.

\bibitem{pena-rodriguezAuAgCore2011}
Peña-Rodríguez,~O.; Pal,~U. Au@ {Ag} core--shell nanoparticles: efficient
  all-plasmonic {Fano}-resonance generators. \emph{Nanoscale} \textbf{2011},
  \emph{3}, 3609--3612.

\bibitem{pena-rodriguezEnhancedFanoResonance2011}
Peña-Rodríguez,~O.; Pal,~U.; Campoy-Quiles,~M.; Rodríguez-Fernández,~L.;
  Garriga,~M.; Alonso,~M. Enhanced {Fano} resonance in asymmetrical {Au}: {Ag}
  heterodimers. \emph{J. Phys. Chem. C} \textbf{2011},
  \emph{115}, 6410--6414.

\bibitem{wangPlasmonicCoreShell2022}
Wang,~H.-J.; Lin,~J.-S.; Zhang,~H.; Zhang,~Y.-J.; Li,~J.-F. Plasmonic
  core--shell materials: synthesis, spectroscopic characterization, and
  photocatalytic applications. \emph{Acc. Mater. Res.}
  \textbf{2022}, \emph{3}, 187--198.

\bibitem{mukherjeeFanoshellsNanoparticlesBuiltin2010}
Mukherjee,~S.; Sobhani,~H.; Lassiter,~J.~B.; Bardhan,~R.; Nordlander,~P.;
  Halas,~N.~J. Fanoshells: nanoparticles with built-in {Fano} resonances.
  \emph{Nano Lett.} \textbf{2010}, \emph{10}, 2694--2701.

\bibitem{chungNanoislandsPlasmonicMaterials2019}
Chung,~T.; Lee,~Y.; Ahn,~M.-S.; Lee,~W.; Bae,~S.-I.; Hwang,~C. S.~H.;
  Jeong,~K.-H. Nanoislands as plasmonic materials. \emph{Nanoscale}
  \textbf{2019}, \emph{11}, 8651--8664.

\bibitem{kvitekPreparationAlloyedCoreshell2020}
Kvitek,~O.; Havelka,~V.; Vesely,~M.; Reznickova,~A.; Svorcik,~V. Preparation of
  alloyed and ``core-shell'' {Au}/{Ag} bimetallic nanostructures on glass
  substrate by solid state dewetting. \emph{J. Alloys Compd.}
  \textbf{2020}, \emph{829}, 154627.

\bibitem{waitkusGoldNanoparticleEnabled2023}
Waitkus,~J.; Chang,~Y.; Liu,~L.; Puttaswamy,~S.~V.; Chung,~T.; Vargas,~A.
  M.~M.; Dollery,~S.~J.; O'Connell,~M.~R.; Cai,~H.; Tobin,~G.~J. Gold
  nanoparticle enabled localized surface plasmon resonance on unique gold
  nanomushroom structures for on-chip {CRISPR}-{Cas13a} sensing.
  \emph{Adv. Mater. Interfaces} \textbf{2023}, \emph{10}, 1--9.

\bibitem{mayergoyz_electrostatic_2005}
Mayergoyz,~I.~D.; Fredkin,~D.~R.; Zhang,~Z. Electrostatic (plasmon) resonances
  in nanoparticles. \emph{Phys. Rev. B} \textbf{2005}, \emph{72},
  155412.

\bibitem{fredkin_resonant_2003}
Fredkin,~D.~R.; Mayergoyz,~I.~D. Resonant behavior of dielectric particles.
  \emph{Phys. Rev. Lett.} \textbf{2003}, \emph{91}, 253902.

\bibitem{mayergoyz_plasmon_2013}
Mayergoyz,~I.~D. \emph{Plasmon {Resonances} in {Nanoparticles}}; WORLD
  SCIENTIFIC, 2013.

\bibitem{forestiere_cloaking_2014}
Forestiere,~C.; Dal~Negro,~L.; Miano,~G. Cloaking of arbitrarily shaped objects
  with homogeneous coatings. \emph{Phys. Rev. B} \textbf{2014}, \emph{89},
  205120.

\bibitem{rakicOpticalPropertiesMetallic1998}
Rakić,~A.~D.; Djurišić,~A.~B.; Elazar,~J.~M.; Majewski,~M.~L. Optical
  properties of metallic films for vertical-cavity optoelectronic devices.
  \emph{Appl. Opt.} \textbf{1998}, \emph{37}, 5271--5283.

\bibitem{xuAuAgCoreShell2025}
Xu,~Z.; Li,~J.; Qi,~J.; Wan,~Y.; Pi,~F. Au@{Ag} core--shell nanoislands
  generated {SERS} sensor for histamine sensitive detection.
  \emph{Microchem. J.} \textbf{2025}, \emph{217}, 114964.

\bibitem{fouadComparativeInsightsStructural2025}
Fouad,~S.; Soliman,~L.; Baradács,~E.; Osman,~N.; Nabil,~M.; Sayed,~M.;
  Tomán,~J.~J.; Mehta,~N.; Erdélyi,~Z. Comparative insights into structural
  and optical properties of {ZnO}/{Ag}/{ZnO} and {Ag}/{ZnO}/{Ag} ternary layer
  thin films. \emph{Nanoscale Adv.} \textbf{2025}, \emph{7},
  7811--7825.

\bibitem{hruskaSituOpticalMonitoring2025}
Hruška,~P.; Jakubik,~M.; More-Chevalier,~J.; Volfova,~L.; Novotný,~M. In situ
  optical monitoring and effect of initial film temperature on pulsed
  laser-induced dewetting of ultrathin {Ag} films. \emph{J. Appl. Phys.} \textbf{2025}, \emph{138}.

\bibitem{diernerInfluenceAuAlloying2024}
Dierner,~M.; Will,~J.; Landes,~M.; Volland,~C.; Branscheid,~R.; Zech,~T.;
  Unruh,~T.; Spiecker,~E. Influence of {Au} alloying on solid state dewetting
  kinetics and texture evolution of {Ag} and {Ni} thin films. \emph{Surf. Interfaces} \textbf{2024}, \emph{46}, 104008.

\bibitem{husainPlasmonicNanoislandFilms2026}
Husain,~S.; Weng,~P.-W.; Yang,~Y.-Y.; Sneka,~C.; Kuo,~T.-R. Plasmonic
  nanoisland films for bacterial theranostics: {A} comprehensive study of
  silver and gold nanoislands for surface-enhanced {Raman} scattering detection
  and photothermal therapy. \emph{FlatChem} \textbf{2026}, 100993.

\bibitem{shermanDistributionSingleParticleResonances2024}
Sherman,~Z.~M.; Milliron,~D.~J.; Truskett,~T.~M. Distribution of
  {Single}-{Particle} {Resonances} {Determines} the {Plasmonic} {Response} of
  {Disordered} {Nanoparticle} {Ensembles}. \emph{ACS Nano} \textbf{2024},
  \emph{18}, 21347--21363.

\bibitem{Fano1961}
Fano,~U. Effects of Configuration Interaction on Intensities and Phase Shifts.
  \emph{Phys. Rev.} \textbf{1961}, \emph{124}, 1866--1878.

\bibitem{maatiExplorationBimetallicAu2022}
Maati,~L. A.~E.; Alkallas,~F.~H.; Trabelsi,~A. B.~G.; Elaissi,~S.;
  Alrebdi,~T.~A.; Ahmad,~M. Exploration of {Bimetallic} {Au}@ {Ag}
  {Core}--{Shell} {Nanocubes} {Dimers} {Supports} {Plasmonic} {Fano}
  {Resonances}. \emph{Plasmonics} \textbf{2022}, \emph{17}, 1843--1855.

\bibitem{nocerinoNextGenerationPlasmonicPlatforms2025}
Nocerino,~V.; Miranda,~B.; Dardano,~P.; Colombelli,~A.; Lospinoso,~D.;
  Manera,~M.~G.; Sanità,~G.; Esposito,~E.; Dello~Iacono,~S.; Rella,~R.
  Next-{Generation} {Plasmonic} {Platforms}: {Hybrid} {Au}--{Si3N4}
  {Nanostructures} for {Scalable} {Sub}-{Femtomolar} {Biosensing}. \emph{ACS Appl. Nano Mater.} \textbf{2025}, \emph{8}, 23690--23700.

\bibitem{mirandaMetalEnhancedFluorescenceImmunosensor2020}
Miranda,~B.; Chu,~K.-Y.; Maffettone,~P.~L.; Shen,~A.~Q.; Funari,~R.
  Metal-{Enhanced} {Fluorescence} {Immunosensor} {Based} on {Plasmonic}
  {Arrays} of {Gold} {Nanoislands} on an {Etched} {Glass} {Substrate}.
  \emph{ACS Appl. Nano Mater.} \textbf{2020}, \emph{3}, 10470--10478.

\bibitem{Bhalla2018}
Bhalla,~N.; Sathish,~S.; Galvin,~C.~J.; Campbell,~R.~A.; Sinha,~A.; Shen,~A.~Q.
  Plasma-assisted large-scale nanoassembly of metal-insulator bioplasmonic
  mushrooms. \emph{ACS Appl. Mater. Interfaces} \textbf{2018},
  \emph{10}, 219--226.

\bibitem{avciTunableFabricationNanoscale2025}
Avci,~M.~B.; Lertvanithphol,~T.; Horprathum,~M.; Tantiwanichapan,~K.;
  Chananonnawathorn,~C.; Hincheeranan,~W.; Yılmaz,~H.~N.; Cetin,~A.~E. Tunable
  {Fabrication} of {Nanoscale} {Structures} via {Solid}-{State} {Thermal}
  {Dewetting} for {Label}-{Free} {Biosensing} {Applications}. \emph{ACS Appl. Nano Mater.} \textbf{2025}, \emph{8}, 15623--15634.

\end{thebibliography}

\end{document}